# Acoustic coupling between cascade sub-chambers and its influence on overall transmission loss

Yuhui Tong

School of Mechanical and Chemical Engineering,

The University of Western Australia, Crawley WA 6009, Australia

Xiang Yu

Institute of High Performance Computing, Agency for Science, Technology and Research, Singapore 138632, Singapore

Jie Pan[a]

School of Mechanical and Chemical Engineering,

The University of Western Australia, Crawley WA 6009, Australia

**Abstract**

In this letter, acoustic interaction between cascade sub-chambers is investigated by modelling the sound field in a silencer with cascade-connected sub-chambers using a sub-structuring technique. The contribution of the acoustic coupling to the net energy flow through each individual sub-chamber is derived quantitatively and the mechanism by which evanescence contributes to the sound transmission loss of the silencer is revealed.

## I. Introduction

Silencers with cascade sub-chambers are commonly used to reduce noise emissions from exhaust and ventilation systems.[1] Since transmission line theory is not capable of dealing with sound transmission of silencers with complex configurations, numerical methods such as the finite element method (FEM)[2] and the boundary element method (BEM)[3] are often used. However, the computational cost of using these numerical methods can be expensive for large system dimensions and optimization tasks where repetitive computations are required. Accordingly, the sub-structuring technique was developed[4,5,6] to overcome this difficulty. It provides flexibility to silencer analysis and design.

Using the sub-structuring technique, a systematic framework for designing a silencer with multiple sub-chambers was proposed by Yu *et al.*[7] Instead of pursuing an optimization for the whole system, their optimizations were conducted respectively for isolated (otherwise cascade-connected) sub-chambers, allowing the effective transmission loss (TL) of each sub-chamber to cover one selected frequency range of interest. They demonstrated that a desired broadband noise attenuation can be obtained by cascade-connecting these sub-chambers. Although their work produced encouraging results for a selected silencer configuration, a challenging question remains on how coupling between the cascade sub-chambers affects the overall TL of the silencer, which motivated this letter.

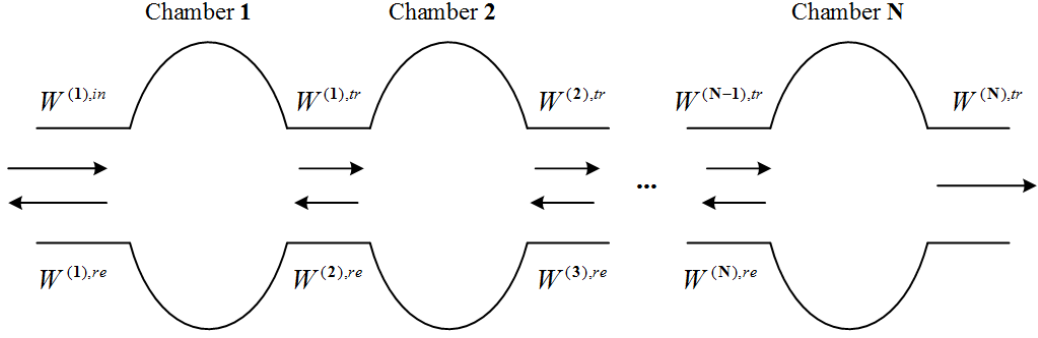

Figure 1 A model silencer with cascade sub-chambers.

Figure 1 shows the schematics of a silencer consisting of cascade sub-chambers. For the silencer, the TL of the **n**$^{th}$ individual sub-chamber is denoted as $TL^{(\mathbf{n})}$, while the actual TL of the **n**$^{th}$ sub-chamber *in situ* is denoted as $TL^{(\mathbf{n}),S}$. While $\sum_n TL^{(\mathbf{n})}$ is used to intuitively estimate the overall noise reduction performance, the exact overall TL of the cascade is instead the sum of the actual TLs of each sub-chamber:

$$TL^{overall} = \sum_{\mathbf{n}}^{N} TL^{(\mathbf{n}),S} = \sum_{\mathbf{n}}^{N} 10\log_{10}(W^{(\mathbf{n}),in}/W^{(\mathbf{n}),tr}), \qquad (1)$$

where $W^{(\mathbf{n}),in}$ and $W^{(\mathbf{n}),tr}$ are respectively the incident and transmitted sound powers of the **n**$^{th}$ sub-chamber within the connected system. When sub-chambers are joined by long connecting ducts, the couplings between them are dominated by propagating waves within the connecting ducts. Additional contributions from evanescent waves need to be taken into account when chambers are closely connected.

## II. Modelling the TL of cascade-connected sub-chambers

### A. Modal scattering matrix of a single chamber

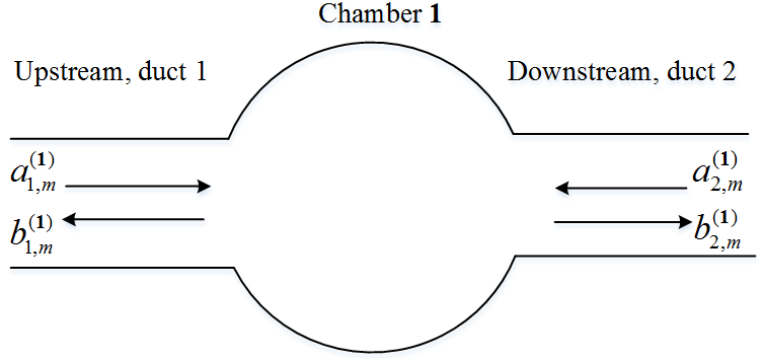

Figure 2 A single chamber connected with two semi-infinite ducts.

Figure 2 shows an acoustic chamber connected to two semi-infinite ducts. The upstream and downstream ducts are denoted as 1 and 2, respectively. The sound pressure in the $q^{th}$ (where $q = 1, 2$) duct in connection with chamber **1** can be written by modal expansion as:

$$p_q^{(1)} = \sum_{m=0}^{\infty} \left( a_{q,m}^{(1)} e^{j\kappa_{q,m}^{(1)} z_q^{(1)}} + b_{q,m}^{(1)} e^{-j\kappa_{q,m}^{(1)} z_q^{(1)}} \right) \psi_{q,m}^{(1)} \left( x_q^{(1)}, y_q^{(1)} \right), \qquad (2)$$

where superscript (**1**) denotes chamber **1**; $(x_q^{(1)}, y_q^{(1)}, z_q^{(1)})$ are the local coordinates, with $z_q^{(1)} = 0$ at the interface between the duct and the chamber and $z_q^{(1)}$ positive away from the chamber; $a_{q,m}^{(1)}$ and $b_{q,m}^{(1)}$ are the amplitudes of the $m^{th}$ incident and scattered (either reflected or transmitted) sound waves; and $\kappa_{q,m}^{(1)} = \sqrt{k^2 - (k_{q,m}^{(1)})^2}$ and $\psi_{q,m}^{(1)}\left(x_q^{(1)}, y_q^{(1)}\right)$ denote respectively the axial wavenumber and the cross-section mode shape of the $m^{th}$ mode in the $q^{th}$ duct of chamber A. It is also worth noting that $q = 1$ corresponds to the upstream duct and $q = 2$ to the downstream duct of the chamber.

Due to the linearity of the system, the complex magnitude of the $m^{th}$ scattered mode in the $q^{th}$ waveguide, $b_{q,m}^{(1)}$, can be expressed as a summation of the scattered waves $b_{q,m,q',m'}^{(1)}$ as a result of incident waves $a_{q',m'}^{(1)}$ ($m' = 0, 1, 2, ...$) from either upstream ($q' = 1$) or downstream ($q' = 2$) ducts, i.e.:

$$b_{q,m}^{(1)} = \sum_{q',m'} b_{q,m,q',m'}^{(1)} = \sum_{q',m'} s_{q,q',m,m'}^{(1)} a_{q',m'}^{(1)}, \qquad (3)$$

where $s_{q,q',m,m'}^{(1)}$ is the modal scattering coefficient that relates the $m^{\text{th}}$ incident mode $a_{q',m'}^{(1)}$ in the $q'^{\text{th}}$ duct and the corresponding $m^{\text{th}}$ scattered mode $b_{q,m,q',m'}^{(1)}$ in the $q^{\text{th}}$ duct. Alternatively, $(\mathbf{R}_{q,q}^{(1)})_{m,m'} = s_{q,q,m,m'}^{(1)}$ and $(\mathbf{T}_{q,q'}^{(1)})_{m,m'} = s_{q,q',m,m'}^{(1)}$ denote respectively the reflection and transmission coefficients of chamber **1**.

## B. Multiple scattering of waves in connecting ducts

The sound transmission of the cascade-connected sub-chambers can be derived using the multiple scattering expansion (MSE) method.[8] Figure 3 shows two sub-chambers connected by a duct. The transmitted wave through the first sub-chamber experiences multiple reflections in the connecting duct between the two sub-chambers. Such multiple reflections were previously discussed in a microwave circuit with two reflecting junctions.[9,10]

Using MSE, for the sound transmission matrix from upstream to downstream, the scattering matrices of the combined system can be expressed by the scattering matrices of each sub-chamber:

$$\mathbf{T}_{2,1}^{(1,2)} = \mathbf{T}_{2,1}^{(2)} \left[ \mathbf{I} - \mathbf{G}_{2,1}^{(1,2)} \mathbf{R}_{2,2}^{(1)} \mathbf{G}_{1,2}^{(1,2)} \mathbf{R}_{1,1}^{(2)} \right]^{-1} \mathbf{G}_{2,1}^{(1,2)} \mathbf{T}_{2,1}^{(1)}, \qquad (4)$$

where $\mathbf{G}_{2,1}^{(1,2)}$ is the propagation matrix[8] for waves propagating in the connecting duct between chamber **1** and chamber **2** from upstream to downstream and $\mathbf{G}_{1,2}^{(1,2)}$ is that in the opposite direction; and $\mathbf{I}$ is the identity matrix. Similarly, other transmission and reflection matrices can be obtained as:

$$\mathbf{T}_{1,2}^{(1,2)} = \mathbf{T}_{1,2}^{(1)}\left[\mathbf{I}-\mathbf{G}_{1,2}^{(1,2)}\mathbf{R}_{1,1}^{(2)}\mathbf{G}_{2,1}^{(1,2)}\mathbf{R}_{2,2}^{(1)}\right]^{-1}\mathbf{G}_{1,2}^{(1,2)}\mathbf{T}_{1,2}^{(2)}, \tag{5}$$

$$\mathbf{R}_{1,1}^{(1,2)} = \mathbf{R}_{1,1}^{(1)} + \mathbf{T}_{1,2}^{(1)}\left[\mathbf{I}-\mathbf{G}_{1,2}^{(1,2)}\mathbf{R}_{1,1}^{(2)}\mathbf{G}_{2,1}^{(1,2)}\mathbf{R}_{2,2}^{(1)}\right]^{-1}\mathbf{G}_{1,2}^{(1,2)}\mathbf{R}_{1,1}^{(2)}\mathbf{G}_{2,1}^{(1,2)}\mathbf{T}_{2,1}^{(1)} \text{ and} \tag{6}$$

$$\mathbf{R}_{2,2}^{(1,2)} = \mathbf{R}_{2,2}^{(2)} + \mathbf{T}_{2,1}^{(2)}\left[\mathbf{I}-\mathbf{G}_{2,1}^{(1,2)}\mathbf{R}_{2,2}^{(1)}\mathbf{G}_{1,2}^{(1,2)}\mathbf{R}_{1,1}^{(2)}\right]^{-1}\mathbf{G}_{2,1}^{(1,2)}\mathbf{R}_{2,2}^{(1)}\mathbf{G}_{1,2}^{(1,2)}\mathbf{T}_{1,2}^{(2)}. \tag{7}$$

For $\mathbf{N}$ cascade-connected sub-chambers, the total scattering matrices can be obtained by applying Eqs. (4)–(7) ($\mathbf{N}-1$) times. Taking $\mathbf{N}=3$ as an example, Eq. (4) becomes

$$\mathbf{T}_{2,1}^{(1,2,3)} = \mathbf{T}_{2,1}^{(3)}\left[\mathbf{I}-\mathbf{G}_{2,1}^{(2,3)}\mathbf{R}_{2,2}^{(1,2)}\mathbf{G}_{1,2}^{(2,3)}\mathbf{R}_{1,1}^{(3)}\right]^{-1}\mathbf{G}_{2,1}^{(2,3)}\mathbf{T}_{2,1}^{(1,2)}. \tag{8}$$

## C. Sound pressure in the connecting ducts

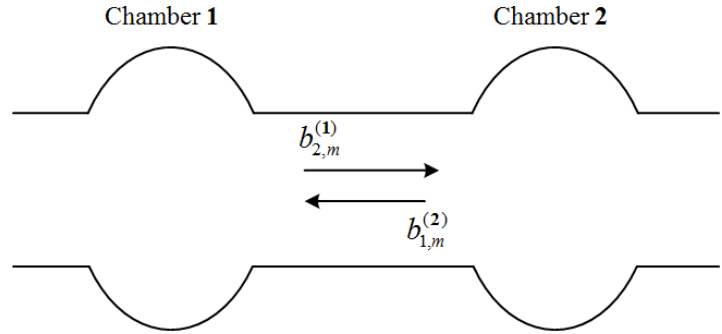

Figure 3 Two sub-chambers joined by a connecting duct.

The sound pressure in the connecting duct in Figure 3 can be written as a superposition of oppositely propagating (or decaying) waves, i.e., $b_{2,m}^{(1)}$ for the amplitude of the $m^{\text{th}}$ mode propagating (or decaying) from upstream to downstream, and $b_{1,m}^{(2)}$ for the amplitude of the $m^{\text{th}}$ mode from downstream to upstream. Applying the MSE technique yields:

$$\mathbf{b}_2^{(1)} = \left\{\left[\mathbf{I}-\mathbf{R}_{2,2}^{(1)}\mathbf{G}_{1,2}^{(1,2)}\mathbf{R}_{1,1}^{(2)}\mathbf{G}_{2,1}^{(1,2)}\right]^{-1}\mathbf{T}_{2,1}^{(1)}\right\}\mathbf{a}_1^{(1)} \text{ and} \tag{9}$$

$$\mathbf{b}_1^{(2)} = \left\{\mathbf{G}_{1,2}^{(1,2)}\left[\mathbf{I}-\mathbf{R}_{1,1}^{(2)}\mathbf{G}_{2,1}^{(1,2)}\mathbf{R}_{2,2}^{(1)}\mathbf{G}_{1,2}^{(1,2)}\right]^{-1}\mathbf{R}_{1,1}^{(2)}\mathbf{G}_{2,1}^{(1,2)}\mathbf{T}_{2,1}^{(1)}\right\}\mathbf{a}_1^{(1)}, \tag{10}$$

where $\mathbf{b}_2^{(1)} = \left[b_{2,0}^{(1)}, b_{2,1}^{(1)}, ...\right]^T$, $\mathbf{b}_1^{(2)} = \left[b_{1,0}^{(2)}, b_{1,1}^{(2)}, ...\right]^T$, and $\mathbf{a}_1^{(1)} = \left[a_{1,0}^{(1)}, a_{1,1}^{(1)}, ...\right]^T$ is the amplitude vector for incident waves towards the first chamber from the upstream duct. Using the sound power expression in the rigid-wall ducts:[13]

$$W = \frac{1}{2} Re\left(\int_S p^* v_z \cdot dS\right), \qquad (11)$$

where $v_z$ and $p$ are the particle velocity (in the $z$-direction) and the sound pressure at the cross-section $S$ of the duct, and these can be determined using Eq. (2) and its derivative with respect to $z$. The incident and transmitted sound powers in the downstream and upstream ducts and in the connecting ducts can be calculated accordingly.

## III. Acoustic coupling between sub-chambers

The acoustic coupling between the cascade-connected sub-chambers is caused by the multiple scattering between them. The multiple scattering is determined by:

(1) The scattering properties of the individual sub-chambers, described by the scattering matrices of the sub-chamber; and

(2) The sound interference properties of the connecting ducts.

The former has been examined extensively,[12] while the effects of propagating and evanescent waves in the connecting ducts on the TL are examined below.

### A. Effect of propagating waves on coupling

If the connecting duct is sufficiently long, then the sound field in the duct will be dominated by the propagating waves, although the pressure distribution in the vicinity of the sub-chambers might be complicated by the contribution of evanescent waves.

Taking coupling between two sub-chambers as an example, the sound power in the connecting duct below the first cut-on frequency is:

$$W^{1,2}(k) = \frac{\kappa_{2,0}^{(1)}}{\rho_0 c_0 k}\left|b_{2,0}^{(1)}\right|^2 - \frac{\kappa_{2,0}^{(1)}}{\rho_0 c_0 k}\left|b_{1,0}^{(2)}\right|^2 = W^{(1),tr} - W^{(2),re}. \quad (12)$$

Recalling the definition of $TL^{(\mathbf{n}),S}$ in Eq. (1), the sound power transmitted between sub-chambers, as well as the actual TLs of each chamber can be obtained by the Eq. (12) so that the overall TL is:

$$TL^{overall} = \sum_{n}^{N} TL^{(\mathbf{n}),S} = \sum_{n=1}^{N} (TL^{(\mathbf{n})} + \Delta TL^{(\mathbf{n})}), \quad (13)$$

where:

$$\Delta TL^{(n)} = TL^{(\mathbf{n}),S} - TL^{(\mathbf{n})} = 20(1-\delta_{\mathbf{n},\mathbf{N}})\log_{10}\left(\left|1 - e^{i2kd_{\mathbf{n},\mathbf{n+1}}} R_{2,2,0,0}^{(\mathbf{n})} R_{1,1,0,0}^{(\mathbf{n+1},\ldots,\mathbf{N})}\right|\right) \quad (14)$$

and $\delta_{i,j}$ is the Kronecker delta. Recall the definition of scattering coefficients: $R_{2,2,0,0}^{(\mathbf{n})}$ is the reflection coefficient of the $\mathbf{n}^{th}$ sub-chamber and $R_{1,1,0,0}^{(\mathbf{n+1},\ldots,\mathbf{N})}$ is the reflection coefficient of all the sub-chambers downstream of the $\mathbf{n}^{th}$ sub-chamber; $\Delta TL^{(\mathbf{n})}$ is the difference between $TL^{(\mathbf{n}),S}$ and $TL^{(\mathbf{n})}$ due to the coupling between the sub-chambers; and $d_{\mathbf{n},\mathbf{n+1}}$ is the length of the connecting duct between the $\mathbf{n}^{th}$ and $(\mathbf{n+1})^{th}$ sub-chambers. A physical interpretation of this difference was qualitatively described in a previous work.[7]

In Eq. (14), if $d_{\mathbf{n},\mathbf{n+1}}$ is much larger than the wavelength, then the term $e^{i2kd_{\mathbf{n},\mathbf{n+1}}}$ changes rapidly with frequency, while $R_{2,2,0,0}^{(\mathbf{n})}$ and $R_{1,1,0,0}^{(\mathbf{n+1},\ldots,\mathbf{N})}$ vary slowly with frequency. The interference between the direct and scattered propagating modes resembles optical multiple-slit interference,[13] and the counterpart of the fringes in the light interference pattern is a series of peaks and dips in the overall TL. Assuming that $\left|R_{2,2,0,0}^{(n)}\right|$ and $\left|R_{1,1,0,0}^{(n+1,...,N)}\right|$ approach 1 (total reflection) in the frequency range of interest, $\Delta TL^{(n)}$ can reach a maximum of 6.02 dB and a minimum of $-\infty$ dB.

Figure 4 shows the TL of a silencer with two sub-chambers, where $\Omega = k/k_1$ is the non-dimensional wavenumber. The parameters are $h_1 = 0.3 \text{ m}$, $w_1 = 0.2$, $h_2 = 0.3$, $w_2 = 0.3$, $\delta_1 = \delta_2 = 0$, and $d_{1,2} = 1.5$. The overall TL calculated by Eq. (1) and using Eqs. (13) and (14) are noted in Fig. 4 as $TL^{overall}$ and $TL^{overall,prop}$, respectively, which agree excellently with each other. Compared to $\sum_n TL^{(n)}$, the overall TL exhibits additional peaks and dips, which as anticipated are the contributions of the interference of propagating waves.

## B. Effect of evanescence on coupling

When cascade sub-chambers are closely connected, evanescent waves cannot be neglected. At some frequencies, evanescent waves can greatly affect the sound power transmission between sub-chambers. As a result, the overall TL will be affected by the evanescent waves.

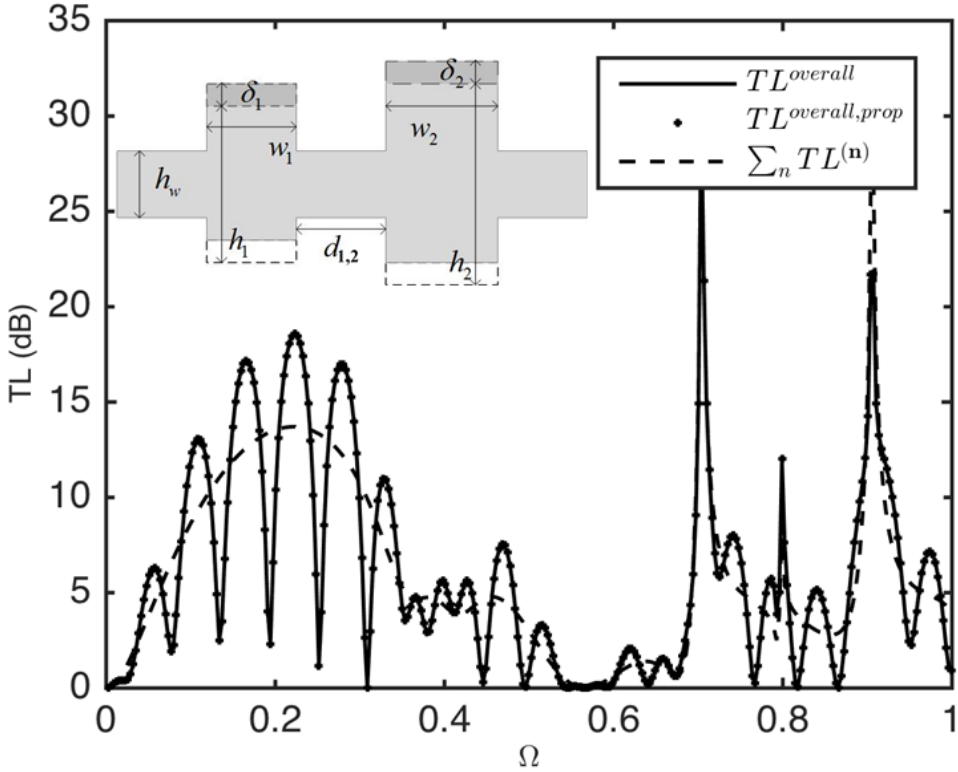

Figure 4 TL of a silencer with two sub-chambers.

Using the silencer in Fig. 4 as an example again, and letting $h_1 = h_2 = 0.3$, $w_1 = w_2 = 0.4$, $\delta_1 = \delta_2 = 0.05$, and $d_{1,2} = 0.02$, the $TL^{overall}$ obtained by Eq. (1) and the $\sum_n TL^{(n)}$ and $TL^{overall,prop}$ obtained by Eqs. (13) and (14) are compared in Fig. 5. The difference between $TL^{overall,prop}$ and $\sum_n TL^{(n)}$ arises from the interference of propagating modes and was indicated in Eq. (14). The $TL^{overall}$ differs from $TL^{overall,prop}$ as a result of taking into account the evanescent waves. Significant differences between $TL^{overall}$ and $TL^{overall,prop}$ are observed at frequencies where contribution of the evanescent waves is important.

The mechanism of transmission of the sound power by evanescent waves is of interest. It is well known that an evanescent wave stores rather than transmits energy. However, a superposition of two oppositely propagating evanescent waves gives rise to a non-zero sound power. For example, the sound power in a connecting duct between two sub-chambers is expressed as:

$$W^{1,2}(k) = \frac{\kappa_{2,0}^{(1)}}{\rho_0 c_0 k}\left|b_{2,0}^{(1)}\right|^2 - \frac{\kappa_{2,0}^{(1)}}{\rho_0 c_0 k}\left|b_{1,0}^{(2)}\right|^2 + \sum_{m=1}^{\infty}\frac{i\kappa_{2,m}^{(1)}}{\rho_0 \omega}\operatorname{Im}[(b_{2,m}^{(1)})^* b_{1,m}^{(2)} - b_{2,m}^{(1)}(b_{1,m}^{(2)})^*]. \tag{15}$$

The last term on the right-hand side of Eq. (15) is the contribution of the evanescent waves in the connecting duct to the total sound power. It is worth noting that the sign of this evanescence term is dependent on the properties of the frequency-dependent wave amplitude.

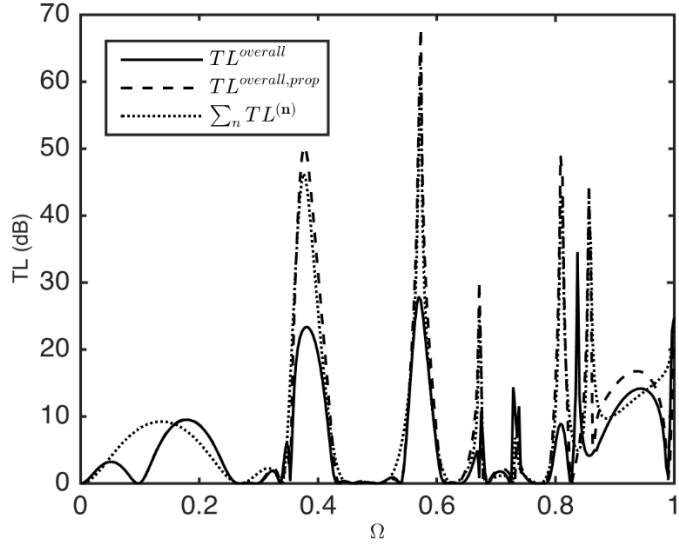

Figure 5 TL of a silencer with two axially asymmetric sub-chambers.

Figure 5 shows that at the frequencies where the evanescent waves are strong, an additional path for sound power transmission via the superposition of evanescent waves from the upstream chamber to the downstream chamber becomes significant, leading to a reduction in TL of up to 40 dB (*e.g.*, at $\Omega = 0.57$). At other frequencies, evanescent waves contribute to additional sound power reflection from the downstream chamber to the upstream chamber, leading to an increase in TL of up to 30 dB (*e.g.*, at $\Omega = 0.83$).

However, the effect of evanescent waves on the overall TL of the silencer is not always strong for closely connected sub-chambers. Recalling Eqs. (6), and (7), the modal coefficients of evanescent waves, *i.e.*, $\mathbf{b}_2^{(1)}$ and $\mathbf{b}_1^{(2)}$ result from the scattering of the incident propagating wave $\mathbf{a}_1^{(1)}$ by the sub-chambers. This means that, in cases where the modal conversion between evanescent and propagating waves is weak, the effect of evanescence can also be weak. Taking the axially symmetric expansion chamber[14] as an example, the amplitudes of evanescent waves, as well as their induced sound power are negligible compared to those of propagating waves when $\Omega<1$. Therefore, the sound power transmission between the connecting chambers is again dominated by the interference of propagating waves. For the axially symmetric counterpart of the first case considered in this sub-section, that is, when $\delta_1=\delta_2=0$, Fig. 6 shows that the effect of evanescence is negligible for most frequencies and that the overall TL is well predicted by Eqs. (13) and (14), in which the effect of evanescent waves was ignored.

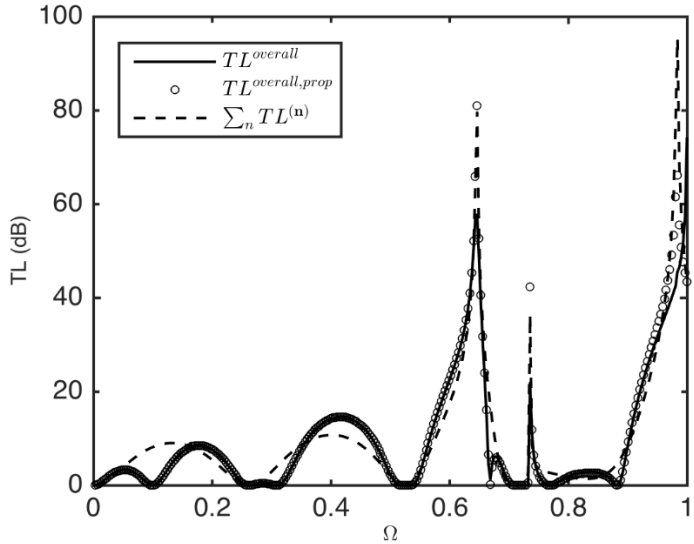

Figure 6 TL of a silencer with two axially symmetric sub-chambers.

# IV. Conclusion

The effects of acoustic interaction between cascade sub-chambers on their sound transmission loss (TL) were investigated. A sound-field silencer with cascade-connected sub-chambers was modelled using the technique of multiple-scattering expansion. The contribution of acoustic coupling to each individual sub-chamber was derived quantitatively by the superposition of waves in the connecting ducts and the mechanism by which evanescence contributes to the transmission of sound power was revealed. Their influences on the total TL of the silencer were examined and discussed.

This study focused on frequencies below the cut-off frequency of the duct and in the absence of a mean flow in the duct. The analysis conducted in this letter can be extended readily to frequencies where higher-order propagating modes are important.

## ACKNOWLEDGEMENTS

The financial support of the Australian Research Council (ARC LP) is gratefully acknowledged.

## References

[1]M. L. Munjal, *Acoustics of Ducts and Mufflers with Application to Exhaust and Ventilation System Design* (John Wiley & Sons, NY, 1987), Chap. 2.

[2]O. Z. Mehdizadeh and M. Paraschivoiu, "A three-dimensional finite element approach for predicting the transmission loss in mufflers and silencers with no mean flow", *Applied Acoustics* **66**(8), 902–918 (2005).

[3]T. W. Wu and G. C. Wan. "Muffler performance studies using a direct mixed-body boundary element method and a three-point method for evaluating transmission loss", *Journal of Vibration and Acoustics* **118**(3), 479–484 (1996).

[4]R. Kirby. "Modeling sound propagation in acoustic waveguides using a hybrid numerical method", *The Journal of the Acoustical Society of America* **124**(4), 1930–1940 (2008).

[5]G. Lou, T. W. Wu, and C. Y. R. Cheng. "Boundary element analysis of packed silencers with a substructuring technique", *Engineering Analysis with Boundary Elements* **27**(7), 643–653 (2003).

[6]M. Ouisse, L. Maxit, C. Cacciolati and J. L. Guyader, "Patch transfer functions as a tool to couple linear acoustic problems", *Journal of Vibration and Acoustics* **127**(5), 458–466 (2005).

[7]X. Yu, Y. Tong, J. Pan and L. Cheng, "Sub-chamber optimization for silencer design", *Journal of Sound and Vibration* **351**, 57–67 (2015).

[8]L. Wirtz, C. Stampfer, S. Rotter, and J. Burgdörfer, "Semiclassical theory for transmission through open billiards: Convergence towards quantum transport", *Physical Review E* **67**(1), 016206 (2003).

[9]R. E. Collin, *Foundations for Microwave Engineering, Second Edition* (Wiley-IEEE Press, NY, 2000), Chap. 5, pp. 347–349.

[10]S. Rotter, F. Libisch, J. Burgdörfer, "Tunable Fano resonances in transport through microwave billiards", *Physical Review E* **69**(4), 046208 (2004).

[11]A. D. Pierce, *Acoustics: An Introduction to its Physical Principles and Applications* (Acoustical Society of America, Melville, NY, 1991), p. 40.

[12]T. Graf, and J. Pan. "Determination of the complex acoustic scattering matrix of a right-angled duct", *The Journal of the Acoustical Society of America* **134**(1) 292–299 (2013).

[13]M. Born and E. Wolf, *Principles of Optics: Electromagnetic Theory of Propagation, Interference and Diffraction of Light* (Cambridge University Press, NY, 1999), Chap. 7.

[14]S. Hein, W. Koch and L. Nannen, "Trapped modes and Fano resonances in two-dimensional acoustical duct–cavity systems", *Journal of Fluid Mechanics* **692**, 257–287 (2012).